\documentclass[]{article}
\usepackage{fullpage}
\usepackage[]{graphicx}
\usepackage{amsmath}
\usepackage{amssymb}
\usepackage{epsfig}
\usepackage{color}
\usepackage{subcaption}
\usepackage{lscape}
\usepackage{epstopdf}
\usepackage[square,numbers]{natbib}
\usepackage {color}
\usepackage{multirow}
\usepackage{setspace}

\usepackage{graphicx}
\usepackage{caption}
\usepackage{subcaption}

 \def\eps{\varepsilon}

 \def\epso{\eps_{\scriptscriptstyle 0}}
\def\lambdao{\lambda_{\scriptscriptstyle 0}}
\def\muo{\mu_{\scriptscriptstyle 0}}
\def\ko{k_{\scriptscriptstyle 0}}

\def\etao{\eta_{\scriptscriptstyle 0}}

\def\##1{{\bf #1}}
\def\=#1{\underline{\underline #1}}
\def\^#1{\breve{#1}}
\def\`#1{{#1^\prime}}
\def\:#1{#1^{\prime\prime}}

\def\les{\left[}
\def\ris{\right]}
\def\lec{\left\{}
\def\ric{\right\}}

\def\r#1{(\ref{#1})}

 \def\epsa{\eps_{\rm a}}
  \def\epsb{\eps_{\rm b}}
   \def\epsc{\eps_{\rm c}}

\def\.{\mbox{ \tiny{$^\bullet$} }}

\def\ux{\hat{\#u}_{\rm x}}
\def\uy{\hat{\#u}_{\rm y}}
\def\uz{\hat{\#u}_{\rm z}}

\def\as{a_{\rm s}}
\def\ap{a_{\rm p}}
\def\rs{r_{\rm s}}
\def\rp{r_{\rm p}}
\def\ts{t_{\rm s}}
\def\tp{t_{\rm p}}

\def\rss{r_{\rm ss}}
\def\rsp{r_{\rm sp}}
\def\rps{r_{\rm ps}}
\def\rpp{r_{\rm pp}}
\def\tss{t_{\rm ss}}
\def\tsp{t_{\rm sp}}
\def\tps{t_{\rm ps}}
\def\tpp{t_{\rm pp}}

\def\Rss{R_{\rm ss}}
\def\Rsp{R_{\rm sp}}
\def\Rps{R_{\rm ps}}
\def\Rpp{R_{\rm pp}}
\def\Tss{T_{\rm ss}}
\def\Tsp{T_{\rm sp}}
\def\Tps{T_{\rm ps}}
\def\Tpp{T_{\rm pp}}

\def\thetainc{\theta_{\rm inc}}

\def\lambdaopBr{\lambda_{\scriptscriptstyle 0_{p_m}}^{Br}}
\def\lambdaosBr{\lambda_{\scriptscriptstyle 0_{s_m}}^{Br}}

\def\lambdaoponeBr{\lambda_{\scriptscriptstyle 0_{p_1}}^{Br}}
\def\lambdaosoneBr{\lambda_{\scriptscriptstyle 0_{s_1}}^{Br}}

\def\lambdaoptwoBr{\lambda_{\scriptscriptstyle 0_{p_2}}^{Br}}
\def\lambdaostwoBr{\lambda_{\scriptscriptstyle 0_{s_2}}^{Br}}

\def\lambdaopthreeBr{\lambda_{\scriptscriptstyle 0_{p_3}}^{Br}}
\def\lambdaosthreeBr{\lambda_{\scriptscriptstyle 0_{s_3}}^{Br}}

\begin{document} 

\begin{center}
{\bf Non-exhibition of Bragg phenomenon by chevronic sculptured thin films: experiment and theory} \\

{Vikas Vepachedu,$^\dag$ Patrick D. McAtee,$^\dag$ and Akhlesh Lakhtakia$^\ast$}\\

{{\rm The Pennsylvania State University, Department of Engineering Science and Mechanics, \\ NanoMM---Nanoengineered Metamaterials Group, 212 EES Building, \\ University Park, PA 16802, USA\\
$^\ast$Corresponding author:  {akhlesh@psu.edu}} \\
$^\dag$These authors contributed equally to this paper.}

\end{center}

\begin{abstract}
The unit cell of a chevronic sculptured thin film (ChevSTF) comprises two identical columnar thin films (CTFs) except that the nanocolumns of the first are oriented at an angle $\chi$ and nanocolumns of the second are oriented at an angle $\pi - \chi$ with respect to the interface of the two CTFs. A ChevSTF containing 10 unit cells  
was fabricated using resistive-heating physical vapor deposition of zinc selenide.  Planewave reflectance and transmittance spectrums of this ChevSTF were measured for a wide variety of incidence conditions over the 500--900-nm range of the free-space wavelength.  Despite its structural periodicity, the ChevSTF did not exhibit the Bragg phenomenon. Theoretical calculations with the CTFs modeled as biaxial dielectric materials indicated that the Bragg phenomenon would not  be manifested  for normal and near-normal incidence, but vestigial manifestation was possible for sufficiently oblique incidence.
Thus, structural periodicity does not always lead to electromagnetic periodicity that underlies the exhibition of the Bragg phenomenon.
 \end{abstract}


\section{Introduction}

The reflection of X~rays of a specific wavelength  incident on a crystal from a specific direction is
intensely peaked in specific directions that are characteristic of the structural periodicity of the
arrangement of atoms in the crystal. A simple relationship describing
this phenomenon is attributed to the father-son duo W.~H. Bragg and W. L. Bragg   \cite{Bragg1,Bragg2}. The underlying mathematical treatment deriving from a rigorous application of the frequency-domain Maxwell equations was provided by
Ewald \cite{Ewald}.

The Bragg phenomenon is exhibited also 
by a periodically nonhomogeneous material, albeit at optical and lower frequencies \cite{Ohtaka,Yab,Maldovan}. Plane waves incident
from any direction in a certain  sector of the full solid angle are highly reflected by a half space
occupied by a periodically nonhomogeneous material,
provided that the free-space wavelength   lies in a certain spectral regime, thereby indicating the formation of partial bandgaps. If high reflectance occurs for all directions in a certain spectral regime, a full bandgap is said to be exhibited. Periodically nonhomogeneous materials have been commonly called as photonic crystals for the last three decades.

The optical response characteristics of one-dimensional photonic crystals in the form of periodic multilayers were investigated in the late 19th century \cite{Rayleigh}. Often called Bragg mirrors or distributed Bragg reflectors nowadays, these structures are fabricated by a variety of thin-film techniques \cite{HWbook,Baumeister,Macleod,Mattox}. The number $N$ of periods must be sufficiently high for the Bragg phenomenon to be exhibited well enough for most applications. 

When a Bragg mirror comprises only isotropic dielectric materials, its performance
is independent of the polarization state of normally incident light. For obliquely incident light,
 the center wavelengths of the high-reflectance bands---called Bragg regimes---do
 not depend on the polarization state of the incident light, but the reflectance
does.

When a Bragg mirror comprises anisotropic dielectric materials, then the 
exhibition of the Bragg phenomenon can be  polarization dependent even for
normally incident light \cite{HWbook,HWapl}. The simplest anisotropic dielectric material
is uniaxial \cite{Chen}, as exemplified by minerals such as calcite, tourmaline, and rutile \cite{Gribble}. Columnar thin films
(CTFs) fabricated using a variety of physical vapor deposition (PVD) techniques   \cite{Macleod, HWbook, STFs} are also anisotropic dielectric materials that are used in polarization-discriminatory Bragg mirrors \cite{HWbook,Baumeister}.  

The unit cell of the simplest Bragg mirror comprises two homogeneous layers of dissimilar materials. The relative permittivity dyadics \cite{Chen} of the two materials must differ in their eigenvalues and/or eigenvectors. But those differences may not be sufficient for the Bragg phenomenon to be exhibited, as has been exemplified by Slepyan and Maksimenko for the following case \cite{NoBP2}. Let a Bragg mirror occupy the region $0 < z < N P$, where $N \gg1$, $P=2L$ is the period of the unit cell, and $L$ is the  thickness of each of the two layers  in the unit cell. Furthermore, let the  relative permittivity dyadics of the  materials in the two layers be denoted by
\begin{equation}
\label{eq1}
\displaystyle{
\left.\begin{array}{l}
\=\eps_1=\=S_{\rm y}(\chi)\.\=\eps^o_{ref}\.\=S_{\rm y}^T(\chi)
\\[5pt]
\=\eps_2=\=S_{\rm y}(\pi-\chi)\.\=\eps^o_{ref}\.\=S_{\rm y}^T(\pi-\chi)
\end{array}\right\}
}
\,,
\end{equation}
where the dyadic
\begin{equation}
\label{eq3}
\=\eps^o_{ref}=\epsa\left(\=I-\ux\ux\right)+\epsb\ux\ux
\end{equation}
contains the eigenvalues
$\epsa$ and $\epsb$  
common to both $\=\eps_1$ and $\=\eps_2$,
$\=I=\ux\ux+\uy\uy+\uz\uz$ is the identity dyadic, the dyadic
\begin{equation}
\=S_{\rm y}(\xi)=\uy\uy +\left(\ux\ux+\uz\uz\right)\cos\xi+\left(\uz\ux-\ux\uz\right)\sin\xi
\end{equation}
indicates a rotation by an angle $\xi$ about the $y$ axis in the $xz$ plane, and the superscript $^T$ denotes the transpose. Equations~\r{eq1} and \r{eq3} indicate that chosen Bragg mirror comprises two   uniaxial dielectric materials that are identical except that
the optical axis
of the first material is oriented at an angle $\chi\in[0,\pi/2)$ with respect to the $+x$ axis 
and the optical axis
of the second material is oriented at an angle $\pi-\chi$ with respect to the $+x$ axis. The optical axes of both materials lie wholly in the $xz$ plane.
Provided that the wave vector of the incident light is also wholly confined to the $xz$ plane, theory shows that no reflection or transmission can occur at the interface of the two materials \cite{NoBP2,NoBP1}, which  implies the absence of the Bragg phenomenon \cite{NoBP2}.

We decided to experimentally verify that theoretical finding of Slepyan and Maksimenko \cite{NoBP2}. For that purpose we fabricated a Bragg mirror made of $N=10$ unit cells.
As this fabrication was accomplished using thermal evaporation \cite{Macleod,STFs}, a PVD technique, each layer
was a CTF with \cite{HWbook,HWapl}
\begin{equation}
\label{eq5}
\=\eps^o_{ref}= \epsb\ux\ux+\epsc\uy\uy+\epsa\uz\uz
\end{equation}
instead of Eq.~\r{eq3}. Thus both layers in the unit cell contained a biaxial dielectric material
with $\epsa\ne\epsb\ne\epsc$. In each period, the nanocolumns of the first CTF are oriented at an angle $\chi$ and nanocolumns of the second CTF are oriented at an angle $\pi - \chi$ with respect to the interface of the two CTFs.
As the unit cell thus has a chevronic morphology, the Bragg mirror
can be classified as a   chevronic sculptured thin film (ChevSTF) \cite{STFs}.
Furthermore, we performed optical characterization experiments with the wave vector of the incident plane wave oriented arbitrarily with respect to the 
$xz$ plane. Thus, our experimental study was  
more general than the theoretical investigation of
Slepyan and Maksimenko \cite{NoBP2}. We also carried out a theoretical  exercise with Eq.~\r{eq5} in order to qualitatively validate our experimental findings.

The plan of this paper is as follows: Section~\ref{methods} describes the experimental and theoretical techniques that we adopted to investigate the optical response characteristics of ChevSTFs, whose cell comprises two CTFs described by
Eqs.~\r{eq1} and \r{eq5}. 
Section~\ref{nrd} presents our experimentally obtained optical data which are
compared qualitatively with the results of our theoretical exercise. Finally, Sec.~\ref{cr}  provides concluding remarks. 
  An $\exp(-i\omega t)$ time-dependence is implicit, with $\omega$ as the angular frequency, $i=\sqrt{-1}$, and $t$ as time. The free-space wavenumber, the
free-space wavelength, and the intrinsic impedance of free space are denoted by $\ko=\omega\sqrt{\epso\muo}$,
$\lambdao=2\pi/\ko$, and
$\etao=\sqrt{\muo/\epso}$, respectively, with $\muo$ and $\epso$ being  the permeability and permittivity of
free space. 


\section{Methods}\label{methods}

\subsection{Reflectances and Transmittances}
The investigated problem can be theoretically formulated as follows. Suppose the region
$0<z<N P$ is occupied by a periodic multilayer of infinite transverse extent. The electric and magnetic
field phasors of the plane wave incident on the periodic multilayer from the half space $z<0$
are given by \cite{STFs}
\begin{eqnarray}
&&\nonumber
\left.\begin{array}{l}
 \#E_{\rm inc}(x,y,z)  = \left(\as  \hat{\#s} + \ap \hat{\#p}_+\right) \, \exp\lec{i   {\ko} \les\left({x\cos\psi+y\sin\psi}\right)\sin\thetainc +z   \cos \thetainc\ris
 }\ric
 \\[5pt]
  \#H_{\rm inc} (x,y,z) = 
 \etao^{-1}\left(\as \hat{\#p}_+ - \ap \hat{\#s}\right)
 \, \exp\lec{i   {\ko} \les\left({x\cos\psi+y\sin\psi}\right)\sin\thetainc +z   \cos \thetainc\ris
 }\ric
 \end{array}\right\}\,,
  \\[5pt]
 &&
 \qquad\qquad\qquad z< 0\,,
 \end{eqnarray} 
 where 
 $\thetainc\in[0^\circ,90^\circ)$ is the angle of incidence with respect to the
 $z$ axis, $\psi\in[0^\circ,360^\circ)$ is the angle of incidence with respect to the
 $x$ axis in the $xy$ plane, $a_s$ is the amplitude of the $s$-polarized component
 of the plane wave, and $a_p$ is the amplitude of the $p$-polarized component
 of the plane wave. Here and hereafter, the unit vectors
 \begin{equation}
 \left.\begin{array}{l}
\hat{\#s} = -\ux \sin\psi + \uy \cos\psi
\\[5pt]
\hat{\#p}_{\pm} = {\mp}(\ux \cos\psi + \uy \sin\psi)\cos\thetainc + \uz \sin\thetainc 
\end{array}\right\}
\end{equation}
delineate the orientations of the field phasors.

The electric and magnetic
field phasors of the reflected plane wave are   given by \cite{STFs}
\begin{eqnarray}
&&
\nonumber
\left.\begin{array}{l}
 \#E_{\rm ref} (x,y,z) = \left(\rs \hat{\#s} + \rp \hat{\#p}_-\right) \, \exp\lec{i   {\ko} \les\left({x\cos\psi+y\sin\psi}\right)\sin\thetainc -z   \cos \thetainc\ris
 }\ric
 \\[5pt]
  \#H_{\rm ref} (x,y,z) = 
 \etao^{-1}\left(\rs \hat{\#p}_- - \rp \hat{\#s}\right)
 \, \exp\lec{i   {\ko} \les\left({x\cos\psi+y\sin\psi}\right)\sin\thetainc -z   \cos \thetainc\ris
 }\ric
 \end{array}\right\}\,,
 \\[5pt]
 &&
 \qquad\qquad\qquad z<0\,,
 \end{eqnarray} 
 and the electric and magnetic
field phasors of the transmitted plane wave   by \cite{STFs}
 \begin{eqnarray}
&&\nonumber
\left.\begin{array}{l}
 \#E_{\rm tr}  (x,y,z)= \left(\ts \hat{\#s} + \tp \hat{\#p}_+\right) \, \exp\lec{i   {\ko} \les\left({x\cos\psi+y\sin\psi}\right)\sin\thetainc +(z-NP)   \cos \thetainc\ris
 }\ric
 \\[5pt]
  \#H_{\rm tr} (x,y,z) = 
 \etao^{-1}\left(\ts \hat{\#p}_+ - \tp \hat{\#s}\right)
 \, \exp\lec{i   {\ko} \les\left({x\cos\psi+y\sin\psi}\right)\sin\thetainc +(z-N P)   \cos \thetainc\ris
 }\ric
 \end{array}\right\}\,,
  \\[5pt]
 &&
 \qquad\qquad\qquad z>N P \,.
 \end{eqnarray} 
 
The reflection amplitudes $\rs$ and $\rp$, as well as the transmission
amplitudes $\ts$ and $\tp$, have to be determined in terms of the
incidence amplitudes $\as$ and $\ap$ in order to find
the reflection and transmission coefficients that appear as the elements of the 
2$\times$2 matrixes in the following
relations:
\begin{equation}
\label{eq10}
\les\begin{array}{c}\rs\\\rp\end{array}\ris
=
\les\begin{array}{cc}\rss & \rsp\\\rps & \rpp\end{array}\ris
\,
\les\begin{array}{c}\as\\\ap\end{array}\ris\,,
\qquad
\les\begin{array}{c}\ts\\\tp\end{array}\ris
=
\les\begin{array}{cc}\tss & \tsp\\\tps & \tpp\end{array}\ris
\,
\les\begin{array}{c}\as\\\ap\end{array}\ris\,.
\end{equation}
Co-polarized coefficients have both subscripts identical, but
cross-polarized coefficients do not. The square of the magnitude
of a reflection or transmission coefficient is the corresponding
reflectance or transmittance;  thus, $\Rsp = \vert\rsp\vert^2$ is
the reflectance corresponding to the reflection coefficient $\rsp$,
and so on.
The principle of conservation of energy mandates
the constraints
$
\Rss + \Rps + \Tss + \Tps \leq 1$ and
$\Rpp + \Rsp + \Tpp + \Tsp \leq 1
$,
the inequalities turning to equalities only in the
absence of dissipation in the region $0<z<N P$. 

\subsection{Theory}
More generally than is needed for a ChevSTF, let us use
\begin{equation}
\label{eq11}
\left.\begin{array}{l}
\=\eps_1=\=S_{\rm y}(\chi_1)\.\left(
\eps_{{\rm b}_1}\ux\ux+\eps_{{\rm c}_1}\uy\uy+\eps_{{\rm a}_1}\uz\uz\right)
\.\=S_{\rm y}^T(\chi_1)
\\[5pt]
\=\eps_2=\=S_{\rm y}(\pi-\chi_2)\.\left(
\eps_{{\rm b}_2}\ux\ux+\eps_{{\rm c}_2}\uy\uy+\eps_{{\rm a}_2}\uz\uz\right)
\.\=S_{\rm y}^T(\pi-\chi_2)
\end{array}\right\}\,
\end{equation}
for a periodic multilayer whose unit cell comprises two different CTFs. Furthermore,
let $L_1$ and $L_2=P-L_1$ be the two layer thicknesses.

After defining the phasors
\begin{equation}
\label{eq12}
\left.\begin{array}{l}
\#e(z)=\#E(x,y,z)  \exp\les{-i   {\ko} \left({x\cos\psi+y\sin\psi}\right)\sin\thetainc }\ris
\\[5pt]
\#h(z)=\#H(x,y,z)  \exp\les{-i   {\ko} \left({x\cos\psi+y\sin\psi}\right)\sin\thetainc }\ris
\end{array}\right\}\,,
\end{equation}
we find that wave propagation in the two CTFs is governed by the 4$\times$4-matrix ordinary differential equations \cite{STFs}
\begin{equation}
\label{eq13}
\displaystyle{
\left.\begin{array}{l}
\frac{d}{dz}\les\#f(z)\ris=i\les\hat{\#p}(\chi_1;\eps_{{\rm a}_1},\eps_{{\rm b}_1},\eps_{{\rm c}_1})\ris\,\les\#f(z)\ris
\\[5pt]
\frac{d}{dz}\les\#f(z)\ris=i\les\hat{\#p}(\pi-\chi_2;\eps_{{\rm a}_2},\eps_{{\rm b}_2},\eps_{{\rm c}_2})\ris\,\les\#f(z)\ris
\end{array}\right\}\,,
}
\end{equation}
where the column 4-vector
\begin{equation}
[\#f(z)] = [\ux\.\#e(z)\quad \uy\.\#e(z)\quad \ux\.\#h(z)\quad \uy\.\#h(z)]^T\,
\end{equation} 
and the 4$\times$4 matrix
\begin{multline}
[\#P(\xi; \epsa,\epsb,\epsc)] = 
\omega 
\displaystyle{\begin{bmatrix}
0 & 0 & 0 & \muo \\
0 & 0 & -\muo & 0 \\
0 & -\epso \epsc & 0 & 0\\
\epso \frac{\epsa \epsb}{\epsa \cos^2 \xi + \epsb \sin^2 \xi} & 0 & 0 \\
\end{bmatrix} }
\\
+ \ko 
\frac{\left(\epsa - \epsb\right)\sin\thetainc}{\epsa \cos^2 \xi + \epsb \sin^2 \xi}
\sin\xi \cos\xi
\displaystyle{\begin{bmatrix}
\cos\psi & 0 & 0 & 0 \\
\sin\psi & 0 & 0 & 0 \\
0 & 0 & 0 & 0\\
0 & 0 & -\sin\psi & \cos\psi \\
\end{bmatrix}} \\+ 
{\omega \muo}  
\frac{\sin^2\thetainc}{\epsa \cos^2 \xi + \epsb \sin^2 \xi}
\displaystyle{\begin{bmatrix}
0 & 0 & \cos\psi \sin\psi & -\cos^2 \psi \\
0 & 0 & \sin^2 \psi & -\cos\psi \sin\psi \\
0 & 0 & 0 & 0\\
0 & 0 & 0 & 0 \\
\end{bmatrix}} \\
+  {\omega \epso} \sin^2\thetainc
\displaystyle{\begin{bmatrix}
0 & 0 & 0 & 0 \\
0 & 0 & 0 & 0 \\
-\cos\psi \sin\psi & \cos^2 \psi & 0 & 0\\
- \sin^2 \psi & \cos\psi \sin\psi & 0 & 0 \\
\end{bmatrix}}\,.
\label{eq14}
\end{multline}
Accordingly,
\begin{equation}
[\#{f}(NP)] = [\#{M}] \, [\#{f}(0)]\,,
\end{equation}
where the 4$\times$4 matrix
\begin{equation}
[\#{M}] = \Big(\exp\lec{i\les\#P(\pi-\chi_2;\eps_{{\rm a}_2},\eps_{{\rm b}_2},\eps_{{\rm c}_2})\ris L_2}\ric\,
\exp\lec{i\les\#P(\chi_1;\eps_{{\rm a}_1},\eps_{{\rm b}_1},\eps_{{\rm c}_1})\ris L_1}\ric
\Big)^N\,.
\label{eq16}
\end{equation}

 Continuity requirements on the tangential components of the electric and magnetic fields across the interfaces $z=0$ and $z=NP$ give rise to \cite{STFs}
\begin{equation}
\begin{bmatrix}
\ts \\
\tp \\
0 \\
0 \\
\end{bmatrix}
= [\#K]^{-1}\, [\#M] \,[\#K]\,
\begin{bmatrix}
\as \\
\ap \\
\rs \\
\rp \\
\end{bmatrix}\,,
\label{eq20}
\end{equation}
where the 4$\times$4 matrix
\begin{equation}
[\#K] =
\begin{bmatrix}
-\sin \psi & -\cos\psi \cos\thetainc & -\sin\psi & \cos\psi \cos\thetainc \\
\cos\psi & -\sin\psi \cos\thetainc & \cos\psi & \sin\psi \cos\thetainc \\
-\etao^{-1} \cos\psi \cos\thetainc &  \etao^{-1} \sin\psi & \etao^{-1} \cos\psi \cos\thetainc & \etao^{-1} \sin\psi\\
-\etao^{-1} \sin\psi \cos\thetainc & -\etao^{-1}\cos\psi & \etao^{-1} \sin\psi \cos\thetainc & -\etao^{-1} \cos\psi \\
\end{bmatrix} \,.
\end{equation}
The solution of Eq.~\r{eq20} provides the four reflection coefficients and
the four  transmission coefficients defined
in Eqs.~\r{eq10}.

When $\psi=0^\circ$ and $0^\circ \leq \thetainc\lesssim 30^\circ$, the center wavelength 
$\lambdaosBr$ of the Bragg regime of order $m$  is predicted to be \cite{HWbook,CBP}
\begin{equation}
\label{eq19-s}
\lambdaosBr= \frac{2}{m}\left(L_1 \sqrt{\eps_{{\rm c}_1}}+L_2 \sqrt{\eps_{{\rm c}_2}}\right)\cos \thetainc, \qquad m\in\lec1,2,3,...\ric \,,
\\
\end{equation}
for $s$-polarized incidence.
Likewise, the center wavelength 
$\lambdaopBr$ of the Bragg regime of order $m$ is predicted to be \cite{HWbook,CBP}
\begin{equation}
\label{eq19-p}
\lambdaopBr= \frac{2}{m}\left(L_1 \sqrt{\eps_{{\rm d}_1}}+L_2 \sqrt{\eps_{{\rm d}_2}}\right)\cos \thetainc \qquad m\in\lec1,2,3,...\ric\, ,
\\
\end{equation}
for $p$-polarized incidence,
where
\begin{equation}
\eps_{{\rm d}_n}=\eps_{{\rm a}_n}\eps_{{\rm b}_n} \left(\eps_{{\rm a}_n}\cos^2\chi_n+\eps_{{\rm b}_n}\sin^2\chi_n\right)^{-1}\,, \qquad n\in\lec1,2\ric\,.
\end{equation}
Equations~\r{eq19-s} and \r{eq19-p} are predicated on the assumption that the constitutive scalars $\eps_{a_n}$, etc., are frequency independent; otherwise,  both equations transform from explicit to implicit equations for the determination of the center wavelengths.

\subsection{Experimental Methods}

\subsubsection{Fabrication of periodic multilayers comprising CTFs}

Each periodic multilayer was fabricated using thermal evaporation \cite{SLml} implemented inside  a low-pressure chamber from Torr International (New Windsor, NY,
USA).  It contains a quartz crystal monitor  (QCM) calibrated to measure the growing thin-film's thickness, a
receptacle to hold the material to be evaporated,  electrodes to
resistively heat the receptacle, and  a substrate holder positioned about 15~cm   above the receptacle.  As the chamber was
customized to grow sculptured thin films \cite{STFs}, it contains two stepper motors to control the rotation of the substrate holder about two mutually orthogonal axes.

$99.995 \%$ pure ZnSe (Alfa Aesar, Ward Hill, MA, USA) was the material of choice due to its high bulk index of refraction and low absorption in the visible spectral regime \cite{transmission_1, transmission_2}, as well as the ease of evaporation. The manufacturer supplied ZnSe  lumps that were crushed into a fine powder. A respirator was worn to avoid the toxic effects of ZnSe on the respiratory system \cite{Alfa}.  Approximately $4.2$~g of ZnSe powder was packed into a tungsten boat (S22-.005W, R. D. Mathis, Long Beach, CA, USA) that served as the receptacle.

A pre-cleaned glass substrate (48300-0025, VWR, Radnor, PA, USA)  was further cleaned in an ethanol bath using an ultrasonicator for $10$~min on each side. After removal from the bath, the substrate was immediately dried with pressurized nitrogen gas. A straight line was marked on the substrate to serve as the $x$ axis. The substrate was secured to the substrate holder using {Kapton\texttrademark} tape (S-14532, Uline, Pleasant Prairie, WI,
USA), and a shutter was interposed between the receptacle and the substrate holder.  By the side of the glass substrate, a silicon wafer was positioned in order to grow a sample for
imaging on a scanning electron microscope.

The  chamber was pumped down to $1$~$\mu$Torr. The current was then slowly increased to $\sim 100$~A, and the shutter was rotated to allow a collimated position of the ZnSe vapor to reach the substrate. The deposition rate as read through the QCM was  manually maintained equal to $0.4\pm 0.02$~nm s$^{-1}$.  After the deposition was complete, the shutter was rotated to prevent further deposition, the current was brought back to 0~A, the chamber was allowed to cool down for 60~min, and was then exposed to the atmosphere.

Each periodic multilayer was chosen to contain $N=10$ unit cells.  This required fabrication to be done in two stages, with 5 unit cells deposited in each stage. By keeping the substrate stationary for  $387$~s for the deposition of each CTF, the value of $P$ was targeted to be $310$~nm.  The substrate was  rotated by $180^\circ$ about a central normal axis passing through it in $0.406$~s 
between the depositions of any two consecutive CTFs. The collimated vapor was directed at an angle $\chi_{v_1}\in(0^\circ,90^\circ]$ with respect to the plane of the substrate to deposit the first CTF in each period, the corresponding angle being $\chi_{v_2}\in(0^\circ,90^\circ]$ for the second CTF in each period.

Two samples were fabricated:
\begin{itemize}
\item
Sample A:
We chose $\chi_{v_1}=20^\circ$   and $\chi_{v_2}=70^\circ$, so that Eqs.~\r{eq11} would hold in  a way  that would allow the exhibition of the Bragg phenomenon. In consequence of $\chi_{v_1}\ne\chi_{v_2}$,
we expected that $L_1{\ne}L_2$ \cite{HWbook}.
More general than a ChevSTF, Sample A should be a Bragg mirror.

\item
Sample B: We chose $\chi_{v_1}=\chi_{v_2}=20^\circ$. The equality of $\chi_{v_1}$ and $\chi_{v_2}$ would ensure that Eqs.~\r{eq1}  and \r{eq5} hold, and we also expected
that $L_1=L_2$ ideally. Thus, Sample B
should be a ChevSTF.

\end{itemize}
Our choices of $  \chi_{v_1}$ and $\chi_{v_2}$ for Samples A and B were guided by  the measured relative permittivity dyadics of CTFs of three different materials \cite{HWbook,Biaxial3}. The QCM tooling factors were $184$ for Samples A and $273$ for
Sample B.

\subsubsection{Optical Characterization}

Transmittance and reflectance measurements  for both Samples A and B were made no more than 24~h after fabrication.  Each sample was kept in a desiccator up until the time of characterization in order to prevent degradation due to moisture adsorption.
 
The experimental setups for reflection and transmission measurements are described in detail elsewhere \cite{Erten}. Briefly,
light from a halogen source (HL-2000, Ocean Optics, Dunedin, FL, USA) was passed through a fiber-optic cable and then through a linear polarizer (GT10, ThorLabs, Newton, NJ, USA); it was either reflected from or transmitted through the sample to be characterized; and was then passed through a second linear polarizer (GT10, ThorLabs) and  a fiber-optic cable to a CCD spectrometer (HRS-BD1-025, Mightex Systems, Pleasanton, CA, USA). 
The transmittances 
$\Tss$, $\Tps$, $\Tpp$, and $\Tsp$ were measured
 for $\thetainc \in [0^{\circ}, \ 70^{\circ}]$ and $\psi\in\lec0^\circ, 90^\circ\ric$, and the reflectances
 $\Rss$, $\Rps$, $\Rpp$, and $\Rsp$  for $\thetainc \in [10^{\circ}, \ 70^{\circ}]$ and $\psi\in\lec0^\circ, 90^\circ\ric$.  
 
 All data were taken in a dark room to avoid noise from external sources. First, the detector measured the intensity $I_{\rm dark}$ when the incident light was switched off,  no sample
was present, and both polarizers had been removed. Then the intensity $I_{\rm s}$ or $I_{\rm p}$ was measured by the detector with no sample present,
 the first polarizer set to make the incident light either $s$ or $p$ polarized, and  the second polarizer  set to pass light of the same linear polarization state.
 Then, the sample was inserted, the second polarizer  was set to make the detected light either $s$ or $p$  polarized, and the intensity $\tilde{I}_{\rm s}$ or $\tilde{I}_{\rm p}$
was recorded. In this way, the measured value of $\Tsp$ was
\begin{equation}
\Tsp=\frac{\tilde{I}_{\rm s}-I_{\rm dark}}{I_{\rm p}-I_{\rm dark}},
\end{equation}
and similarly for the other seven  reflectances and  transmittances. These measurements were
made for $\lambdao\in[500,900]$~nm.  

\subsubsection{Morphological Characterization}
The cross-sectional morphologies of both Samples A and B  were characterized using an FEI Nova\textsuperscript{TM} NanoSEM 630 (FEI, Hillsboro, OR, USA) field-emission scanning electron microscope.  To get a clear image of morphology  away from any edge-growth effects, each sample was cleaved using the freeze-fracture technique.
This technique is delicate enough to preserve biological samples \cite{Freeze_Fracture}. All samples were sputtered with iridium using a Quorum Emitech\textsuperscript{\copyright} K575X (Quorum Technologies, Ashford, Kent, United Kingdom) sputter coater before imaging.

\section{Results and Discussion}\label{nrd}

\subsection{Morphology}\label{morph-res}

\begin{figure}[ht]
        \centering
        \includegraphics[scale=0.35]{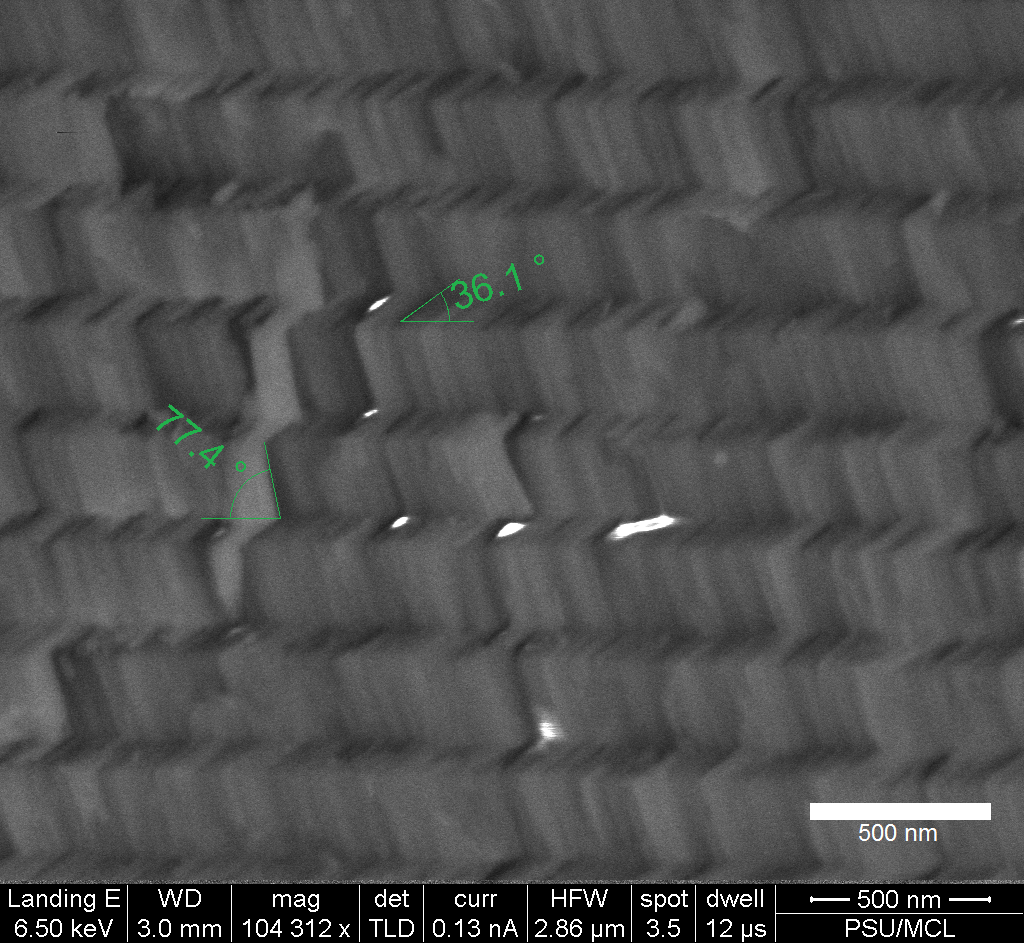}
         \caption{Cross-sectional scanning-electron micrograph of  Sample A (fabricated with $\chi_{v_1}=20^\circ$   and $\chi_{v_2}=70^\circ$) on a silicon wafer.
         \label{Fig1}}
\end{figure}

Figure~\ref{Fig1} presents a cross-sectional scanning-electron micrograph of  Sample A fabricated by setting $\chi_{v_1}=20^\circ$ and $\chi_{v_2}=70^\circ$. Eight of the 10 unit cells are clearly evident with  {$P\simeq313$~nm}. Whereas the nanocolumns in one of the two CTFs in a unit cell are inclined at $36.1^\circ$ with respect to the $+x$ axis, those in the other CTF are
inclined at $102.6^\circ$ with respect to the $+x$ axis. Thus,  the
conditions set forth as Eqs.~\r{eq11} are fulfilled with $\chi_1\simeq 36.1^\circ$ and $\chi_2\simeq 77.4^\circ$. Furthermore, $L_1 \simeq 0.25 P$ whereas  $L_2 \simeq 0.75  P$,
because $\chi_{v_1}$ differs from $\chi_{v_2}$.  Sample A is clearly not a ChevSTF. 
Parenthetically, the inequalities $\chi_1>\chi_{v_1}$ and $\chi_2>\chi_{v_2}$ are in consonance with a plethora of morphological data on CTFs \cite{HWbook,Messier97}.

\begin{figure}[ht]
        \centering
        \includegraphics[scale=0.45]{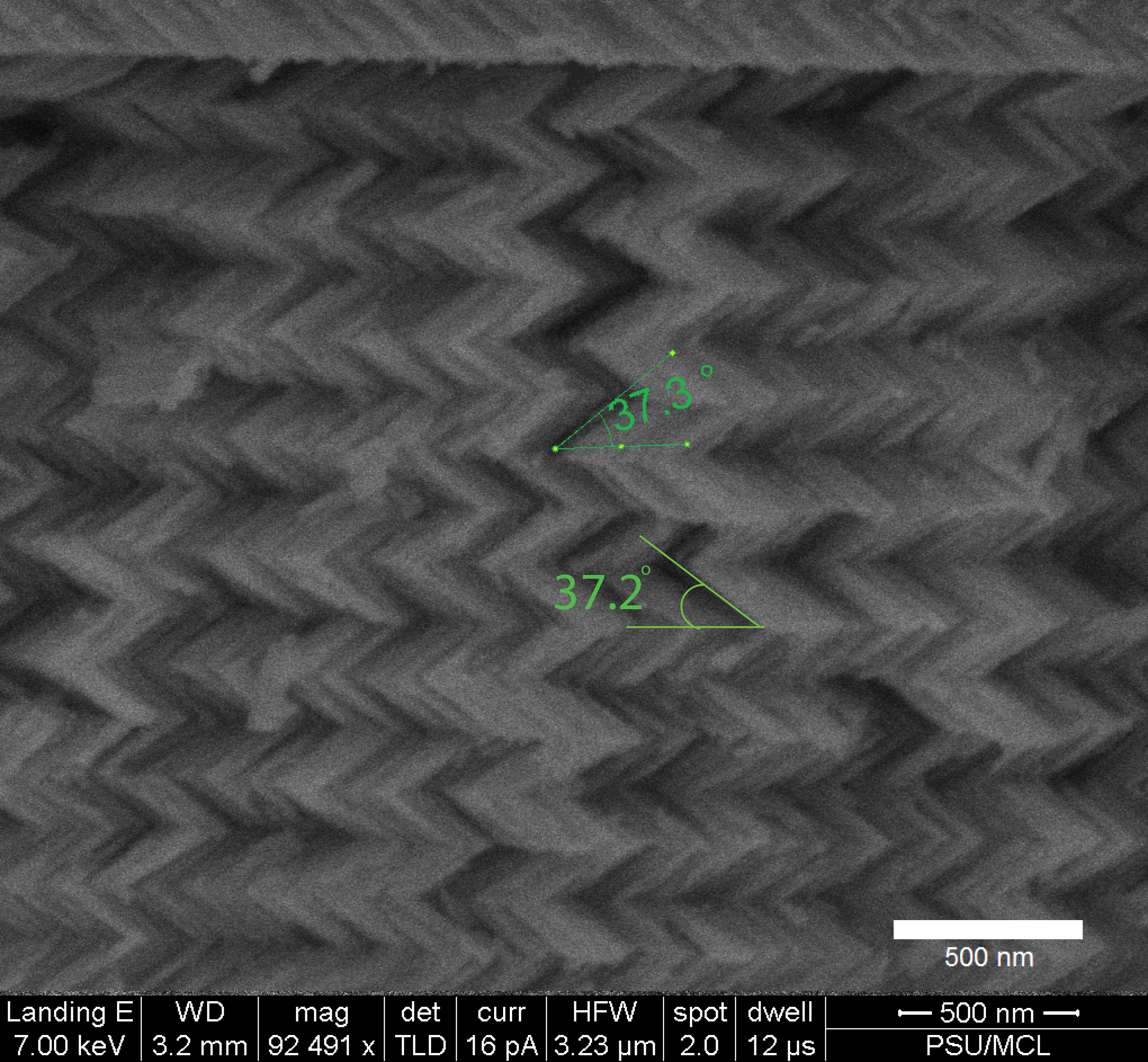}
         \caption{Cross-sectional scanning-electron micrograph of  Sample B (fabricated with $\chi_{v_1}=\chi_{v_2}=20^\circ$) on a silicon wafer.
         \label{Fig2}}
\end{figure}

Figure~\ref{Fig2} presents a cross-sectional scanning-electron micrograph of  Sample B fabricated by setting $\chi_{v_1}=\chi_{v_2}=20^\circ$. All 10 unit cells of this sample are clearly evident with  {$P\simeq 308$~nm} and $L_1 \simeq L_2 \simeq P/2$. Whereas the nanocolumns in one of the two CTFs in a unit cell are inclined at $37.3^\circ$ with respect to the $+x$ axis, those in the other CTF are
inclined at $142.8^\circ$ with respect to the $+x$ axis.  Thus,  the
conditions set forth as Eqs.~\r{eq1} and \r{eq5} are fulfilled with $\chi\simeq 37.25^\circ$, and Sample B is  a ChevSTF.

\subsection{Theoretical Results}\label{theory-res}

Experimental data on the relative permittivity dyadics of CTFs fabricated by evaporating ZnSe are unavailable, but  the  data on the relative permittivity dyadics of CTFs fabricated by evaporation of TiO$_2$ are available \cite{Biaxial3}. The bulk refractive index of ZnSe is $2.5435$ at $\lambdao=700$~nm \cite{Marple}, whereas that of
TiO$_2$ is $2.5512$ 
at the same wavelength  \cite{Devore}. The closeness of the two bulk refractive indexes allowed us to substitute ZnSe by TiO$_2$ for theoretical calculations that can be qualitatively compared with the experimental results.

For $\chi_{v_1}=20^\circ$, data derived from optical-characterization experiments on CTFs of TiO$_2$ 
at $\lambdao=633$~nm \cite{Biaxial3}
are as follows:
$\chi_1=46.367^\circ$, $\eps_{{\rm a}_1}=2.5135$
 $\eps_{{\rm b}_1}=3.9428$,  and $\eps_{{\rm c}_1}=3.1525$. Likewise, for
 $\chi_{v_2}=70^\circ$, data derived from the same optical experiments are as follows:
$\chi_2=82.802^\circ$, $\eps_{{\rm a}_2}=5.5054$,
 $\eps_{{\rm b}_2}=5.8581$,  and $\eps_{{\rm c}_2}=5.5785$.
 For a Bragg mirror comprising TiO$_2$ CTFs with $\chi_{v_1}=20^\circ$, $\chi_{v_2}=70^\circ$,
$L_1=77.5$~nm, and $L_2=232.5$~nm
 (similar to Sample~A),  Eqs.~\r{eq19-s} and \r{eq19-p} yield  $\lambdaostwoBr=686.74$~nm and  $\lambdaoptwoBr=680.87$~nm when $\thetainc=\psi=0^\circ$.
For a ChevSTF comprising TiO$_2$ CTFs with $\chi_{v_1}=\chi_{v_2}=20^\circ$
and $L_1=L_2=155$~nm
(similar to Sample~B), $\lambdaostwoBr=550.41$~nm and  $\lambdaoptwoBr=540.32$~nm when $\thetainc=\psi=0^\circ$, according to
Eqs.~\r{eq19-s} and \r{eq19-p}.

\subsubsection{Bragg mirror}\label{theo-BM}
\begin{figure}[ht]
        \centering
        \includegraphics[scale=0.18]{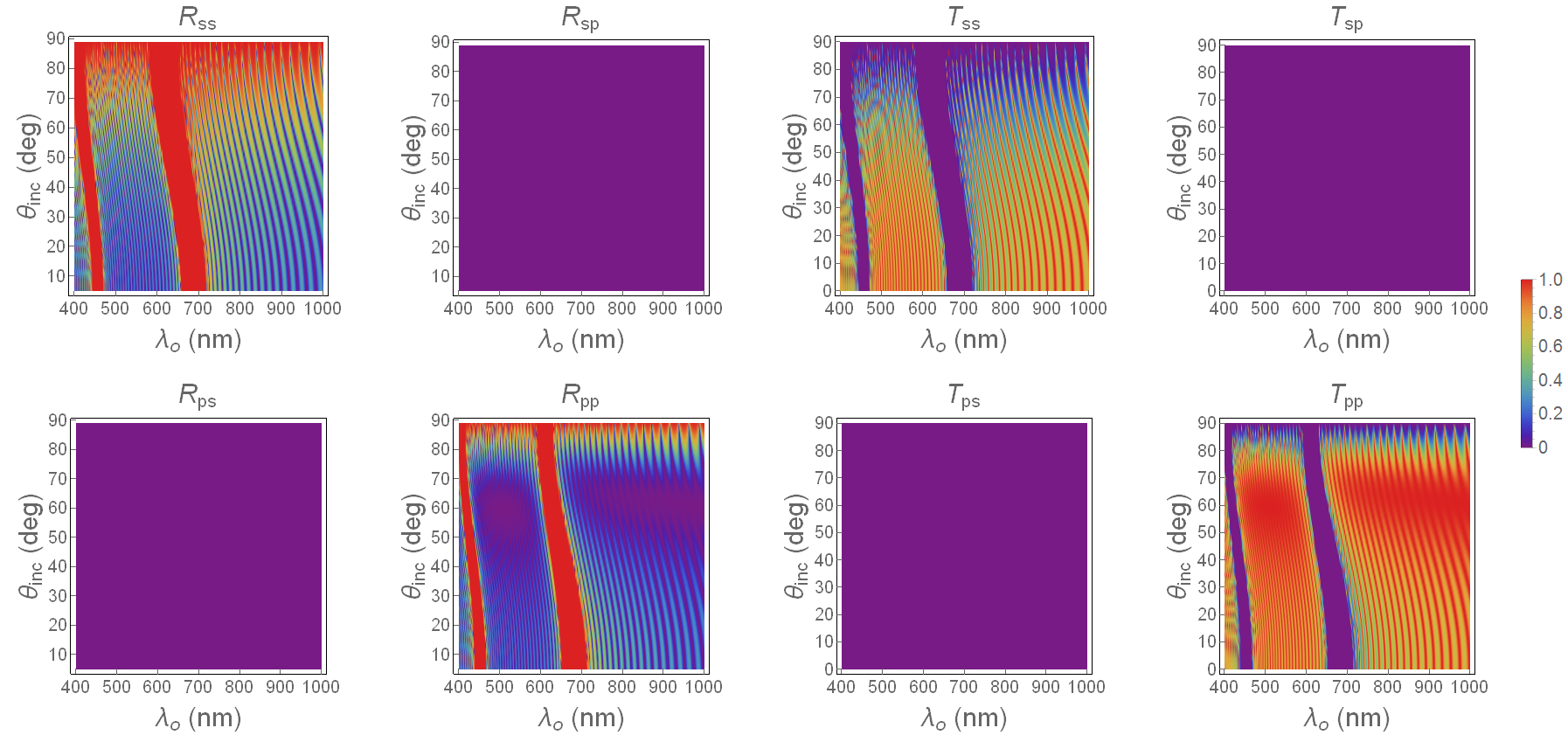}
         \caption{Density plots of all four reflectances and all four transmittances calculated as functions
         of $\lambdao$ and $\thetainc$  for
         a Bragg mirror comprising TiO$_2$ CTFs with $\chi_{v_1}=20^\circ$, $\chi_{v_2}=70^\circ$,
$L_1=77.5$~nm, and $L_2=232.5$~nm (similar to Sample~A),
         when $\psi=0^\circ$. All calculations were made with $N=30$.
         \label{Fig3}}
\end{figure}

\begin{figure}[ht]
        \centering
        \includegraphics[scale=0.18]{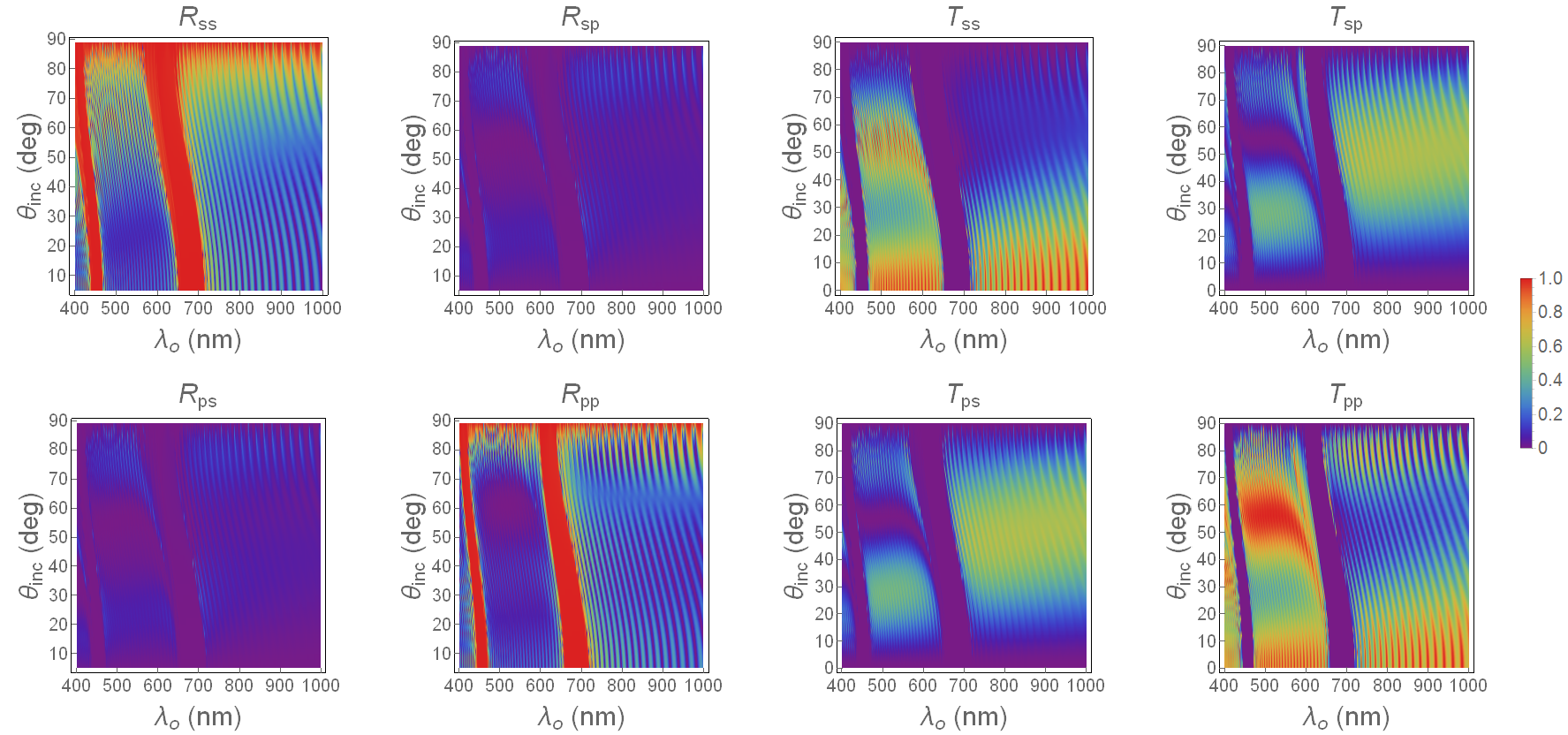}
         \caption{Same as Fig.~\ref{Fig3}, except that
          $\psi=90^\circ$. 
         \label{Fig4}}
\end{figure}

In order to present the characteristic features of the Bragg phenomenon, let us begin with the spectrums of all eight remittances calculated for
the Bragg mirror comprising TiO$_2$ CTFs $\chi_{v_1}=20^\circ$, $\chi_{v_2}=70^\circ$,
$L_1=77.5$~nm, and $L_2=232.5$~nm
(corresponding to Sample A).  These remittances are presented as functions
of $\lambdao\in[400,1000]$~nm and $\thetainc\in[0^\circ,90^\circ)$ in Fig.~\ref{Fig3} for
$\psi=0^\circ$ and in Fig.~\ref{Fig4}  for
$\psi=90^\circ$. For all calculations we set $N=30$ in order to ensure that
 the Bragg phenomenon is manifested well \cite{Baumeister,StJohn}.

Evidence of the Bragg phenomenon of order $m\in\left\{2,3\right\}$ is easy to find
in  Fig.~\ref{Fig3}. This becomes
clear on comparison with the predictions
$\lambdaostwoBr=686.74$~nm, $\lambdaoptwoBr=680.87$~nm, 
$\lambdaosthreeBr=457.83$~nm, and $\lambdaopthreeBr=453.91$~nm, 
of  Eqs.~\r{eq19-s} and \r{eq19-p} for $\thetainc=\psi=0^\circ$. The high-reflectance
band of order $m\in\lec2,3\ric$ in the plot of $\Rss$ is slightly redshifted
from the high-reflectance
band of order $m\in\lec2,3\ric$ in the plot of $\Rpp$. Furthermore, the
Bragg regime of order $m\in\lec2,3\ric$ is blueshifted
with increase in $\thetainc$. Both high-reflectance bands are mirrored
by low-transmittance bands in the plots of $\Tss$ and $\Tpp$. 

With
$\lambdaosoneBr=1373.48$~nm and $\lambdaoponeBr=1361.74$~nm predicted
by  Eqs.~\r{eq19-s} and \r{eq19-p}, evidence of the Bragg phenomenon
of order $m=1$ is  absent from Fig.~\ref{Fig3}. This is not surprising because
the plots in this figure were drawn for only $\lambdao\in[400,1000]$~nm.
By direct computation,
we have verified the existence of high-reflectance bands in the plots
of $\Rss$ and $\Rpp$, as well as of low-transmittance bands in the
plots of $\Tss$ and $\Tpp$, that signify the existence of the Bragg
phenomenon for $m=1$.

Although no convenient formula is available to predict the center wavelengths for
$\psi\in(0^\circ,180^\circ)$, comparison of Figs.~\ref{Fig3} and \ref{Fig4} suffices
to convince that the latter also provides evidence of the Bragg phenomenon of order $m\in\left\{2,3\right\}$. The spectral characteristics of the high-reflectance bands
in the plots of $\Rss$ and $\Rpp$ in Fig.~\ref{Fig4} are the same as in Fig.~\ref{Fig3}.
We found similar characteristics for several values of $\psi\notin\lec0^\circ,90^\circ\ric$
(data not presented here).

\subsubsection{Chevronic STF}\label{theo-ChevSTF}

\begin{figure}[ht]
        \centering
        \includegraphics[scale=0.18]{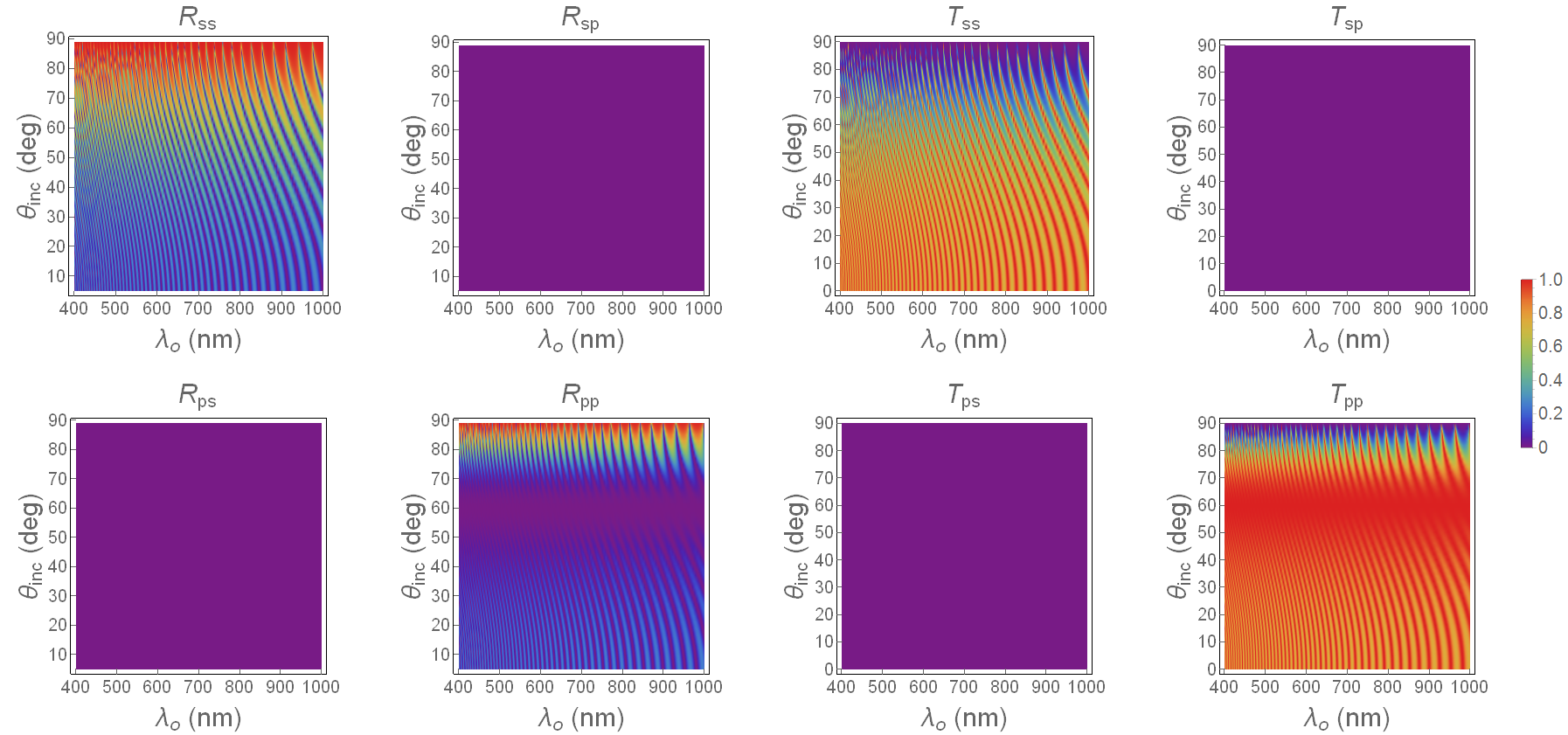}
         \caption{Density plots of all four reflectances and all four transmittances calculated as functions
         of $\lambdao$ and $\thetainc$  for
         a ChevSTF comprising TiO$_2$ CTFs with $\chi_{v_1}=\chi_{v_2}=20^\circ$ 
         and $L_1=L_2=155$~nm (similar to Sample~B),
         when $\psi=0^\circ$. All calculations were made with $N= 30$.
         \label{Fig5}}
\end{figure}

\begin{figure}[ht]
        \centering
        \includegraphics[scale=0.30]{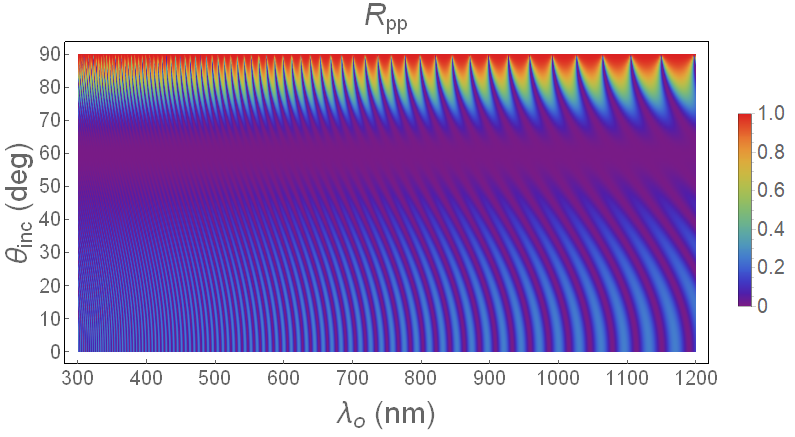}
         \caption{Density plot of $\Rpp$ calculated as a function
         of $\lambdao$ and $\thetainc$  for
         a ChevSTF comprising TiO$_2$ CTFs with $\chi_{v_1}=\chi_{v_2}=20^\circ$ 
         and $L_1=L_2=155$~nm (similar to Sample~B),
         when $\psi=0^\circ$.          \label{Fig6}}
\end{figure}

Next, let us present the spectrums of all eight remittances calculated for a ChevSTF comprising TiO$_2$ CTFs with $\chi_{v_1}=\chi_{v_2}=20^\circ$ and $L_1=L_2=155$~nm
(similar to Sample B). Figure~\ref{Fig5} presents these remittances as functions
of $\lambdao\in[400,1000]$~nm and $\thetainc\in[0^\circ,90^\circ)$ when $\psi=0^\circ$.
Although
$\lambdaostwoBr=550.41$~nm and $\lambdaoptwoBr=540.32$~nm when $\thetainc=\psi=0^\circ$, according to
Eqs.~\r{eq19-s} and \r{eq19-p}, no evidence of the Bragg phenomenon 
of order $m=2$ exists in  Fig.~\ref{Fig5}. Furthermore, although
$\lambdaosoneBr=1100.82$~nm,
 $\lambdaoponeBr=1080.64$~nm, $\lambdaosthreeBr=366.94$~nm, and
 $\lambdaopthreeBr=360.21$~nm are predicted by Eqs.~\r{eq19-s} and \r{eq19-p}
 for  $\thetainc=\psi=0^\circ$, we have verified by direct calculation
 that the Bragg phenomenon of order $m\in\lec1,3\ric$  is  also not manifested 
 for the chosen chevronic STF. This is evident from the plot of $\Rpp$
 in Fig.~\ref{Fig6} for $\lambdao\in[400,1200]$~nm and $\thetainc\in[0^\circ,90^\circ)$.

\begin{figure}[ht]
        \centering
        \includegraphics[scale=0.18]{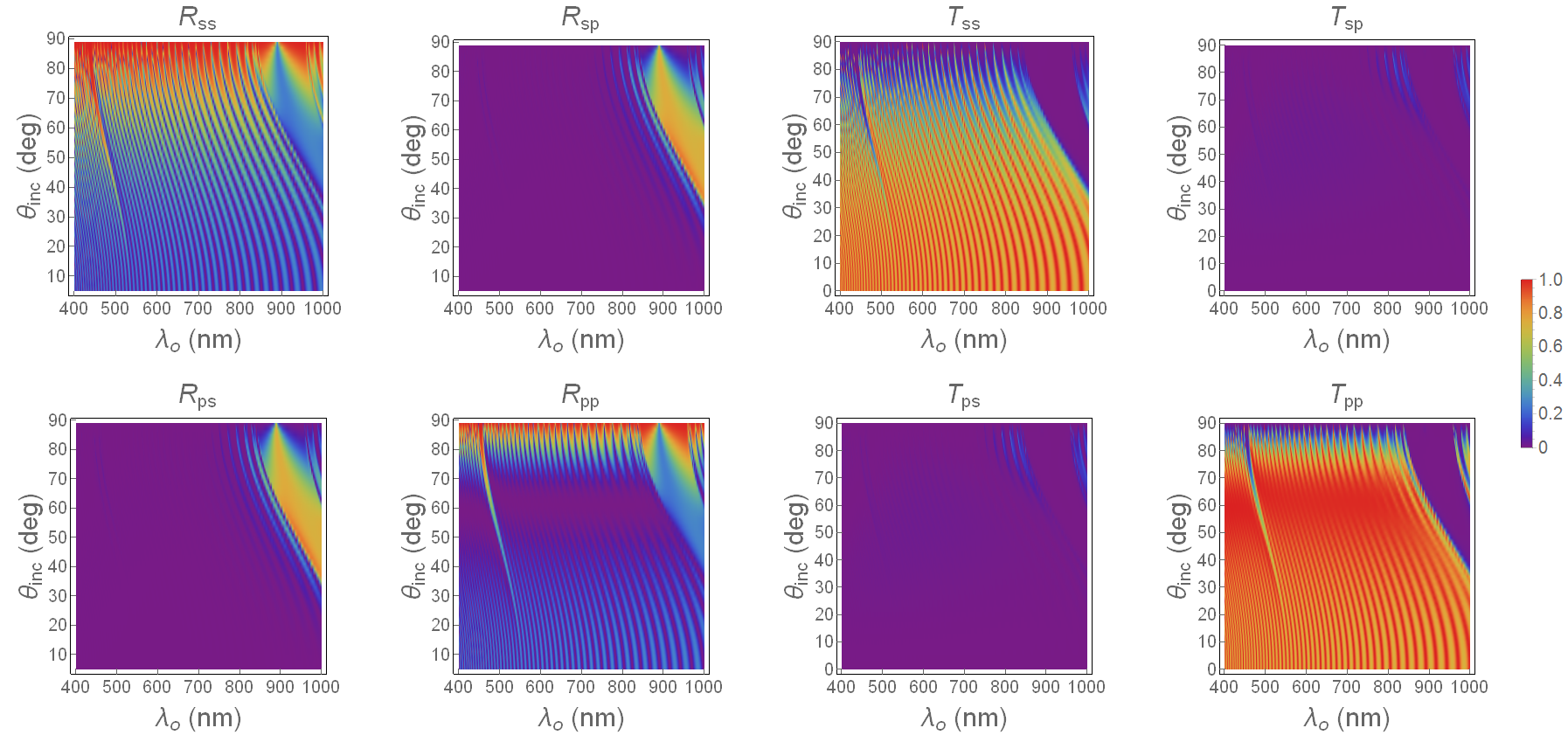}
         \caption{Same as Fig.~\ref{Fig5}, except that
          $\psi=90^\circ$. 
         \label{Fig7}}
\end{figure}

\begin{figure}[ht]
        \centering
        \includegraphics[scale=0.30]{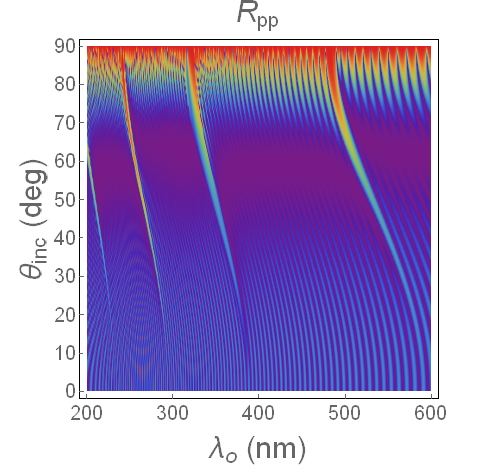}
         \caption{Same as Fig.~\ref{Fig6}, except that
          $\psi=90^\circ$. 
         \label{Fig8}}
\end{figure}

For $\psi=90^\circ$, the situation changes somewhat. In Fig.~\ref{Fig7}, the Bragg phenomenon
of order $m=2$ is  present in the plots of $\Rss$, $\Rpp$, $\Tss$, and $\Tpp$. But this existence
is vestigial at best, as it is evident in very narrow spectral regimes and that too only for $\thetainc\gtrsim40^\circ$. The same characteristic holds true for the Bragg phenomenon of order $m\in\lec3,4,5\ric$, as exemplified by Fig.~\ref{Fig8}.

The situation is somewhat different for the Bragg phenomenon of order $m=1$. High-reflectance bands exist in the plots of $\Rps$ and $\Rss$ for $\thetainc\gtrsim 20^\circ$, indicating that reflection is accompanied by a rotation of the vibration ellipse by angles close to $\pm 85^\circ$.
In other words, the Bragg phenomenon is accompanied by significant polarization conversion,
but even that is not manifested for normal and near-normal incidences.

The foregoing conclusions from examination of computed data can be explained through Eqs.~\r{eq14} and \r{eq16}. For normal incidence, the matrix $[\#P(\xi; \epsa,\epsb,\epsc)]$ simplifies to yield
\begin{eqnarray}
&&
\nonumber
[\#P(\chi; \epsa,\epsb,\epsc)] \Bigg\vert_{\thetainc=0}
= [\#P(\pi-\chi; \epsa,\epsb,\epsc)]\Bigg\vert_{\thetainc=0}
\\[5pt]
&&\qquad = 
\omega 
\displaystyle{\begin{bmatrix}
0 & 0 & 0 & \muo \\
0 & 0 & -\muo & 0 \\
0 & -\epso \epsc & 0 & 0\\
\epso \frac{\epsa \epsb}{\epsa \cos^2 \chi + \epsb \sin^2 \chi} & 0 & 0 \\
\end{bmatrix} }\,.
\label{eq23}
\end{eqnarray}
Then, both CTFs in the unit cell of a ChevSTF are electromagnetically identical and the structural periodicity of the ChevSTF does not translate into electromagnetic periodicity.
Accordingly, the Bragg phenomenon cannot be exhibited for $\thetainc=0^\circ$. 

As $\thetainc$ increases, first the second term and then the third and fourth terms
on the right side of Eq.~\r{eq14} become increasingly consequential. Therefore, $[\#P(\chi; \epsa,\epsb,\epsc)]$ begins to differ from $[\hat{\#p}(\pi-\chi; \epsa,\epsb,\epsc)]$, but the difference 
\begin{eqnarray}
\nonumber
&&[\#P(\chi; \epsa,\epsb,\epsc)]-[\#P(\pi-\chi; \epsa,\epsb,\epsc)]
\\[5pt]
&&\qquad=
2 \ko 
\frac{\left(\epsa - \epsb\right)\sin\thetainc}{\epsa \cos^2 \chi + \epsb \sin^2 \chi}
\sin\chi \cos\chi
\displaystyle{\begin{bmatrix}
\cos\psi & 0 & 0 & 0 \\
\sin\psi & 0 & 0 & 0 \\
0 & 0 & 0 & 0\\
0 & 0 & -\sin\psi & \cos\psi \\
\end{bmatrix}}
\label{eq24}
\end{eqnarray}
stems 
only from the second term on the right side of Eq.~\r{eq14}. Hence,  the Bragg phenomenon would very weakly exhibited for near-normal
incidence. For more appreciable exhibition, $\thetainc$ would have to exceed some threshold value that depends not only on the values of $\eps_{a,b,c}$ and $\chi$ but also of $\psi$.
Equation~\r{eq14} also indicates that
polarization conversion  is not possible when $\sin\psi=0$ \cite{STFs}.

\subsection{Experimental Results}\label{exp-res}

\subsubsection{Bragg mirror}\label{expt-BM}

\begin{figure}[ht]
        \centering
        \includegraphics[scale=0.35]{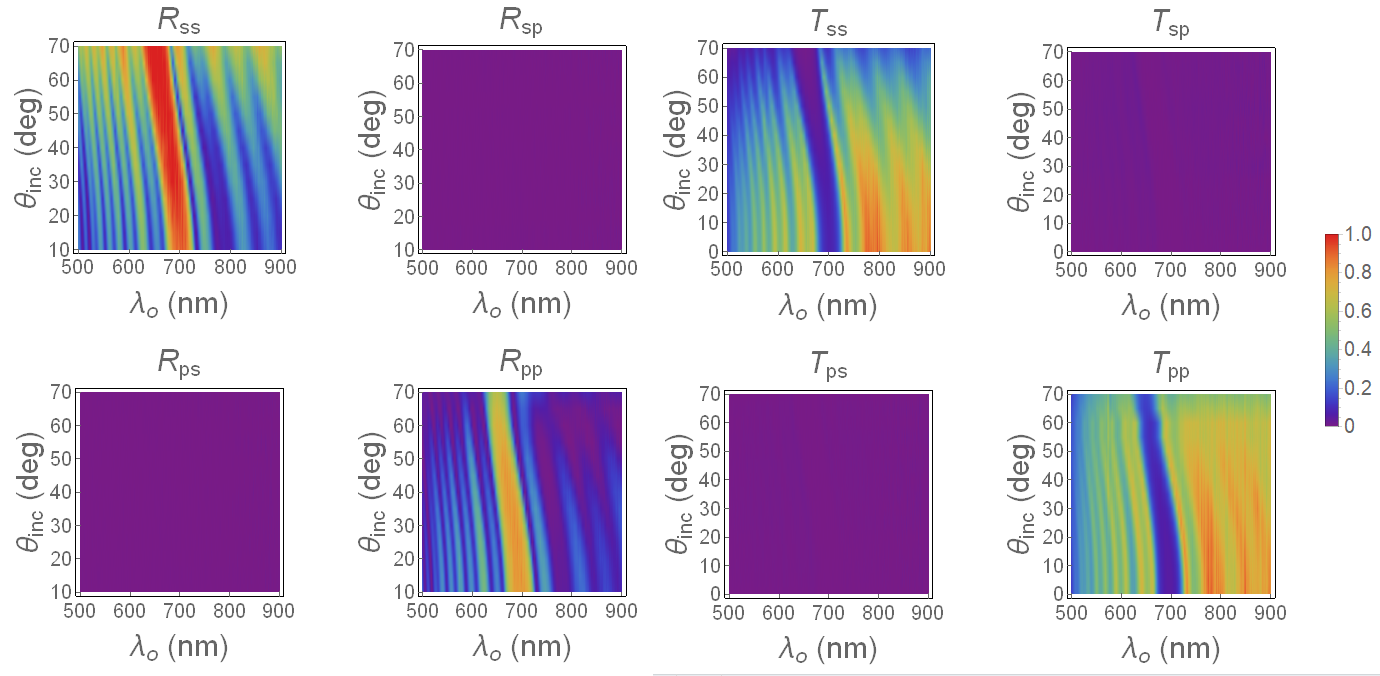}
         \caption{Density plots of all four reflectances and all four transmittances measured as functions
         of $\lambdao$ and $\thetainc$  for Sample A (see Fig.~\ref{Fig1}) when  $\psi=0^\circ$.
          \label{Fig9}}
\end{figure}

Following Sec.~\ref{theory-res},    let us begin with the measured spectrums  
of all eight remittances of Sample~A, which is expected to perform as a Bragg mirror. Figure~\ref{Fig9} presents the reflectances as functions
of $\thetainc\in[10^\circ,70^\circ]$ and transmittances as  functions
of $\thetainc\in[0^\circ,70^\circ]$ for $\lambdao\in[500,900]$~nm  
when $\psi=0^\circ$. As predicted in Sec.~\ref{theory-res}, the Bragg phenomenon
of order $m=2$ is evident as a high-reflectance band each in the plots of $\Rss$ and $\Rpp$
and as a low-transmittance band each in the plots of $\Tss$ and $\Tpp$. Let us note $\Rss$ exceeds $\Rpp$ in the high-reflectance band, especially as $\thetainc$ increases.
The center wavelengths $\lambdaostwoBr$ and $\lambdaoptwoBr$ for $\thetainc=0^\circ$
nearly overlap at approximately $700$~nm and $695$~nm, respectively, in the
plots of $\Tss$ and $\Tpp$.
The Bragg phenomenon of order $m\ne 2$ lies outside the spectral range of our apparatus.

\begin{figure}[ht]
        \centering
        \includegraphics[scale=0.35]{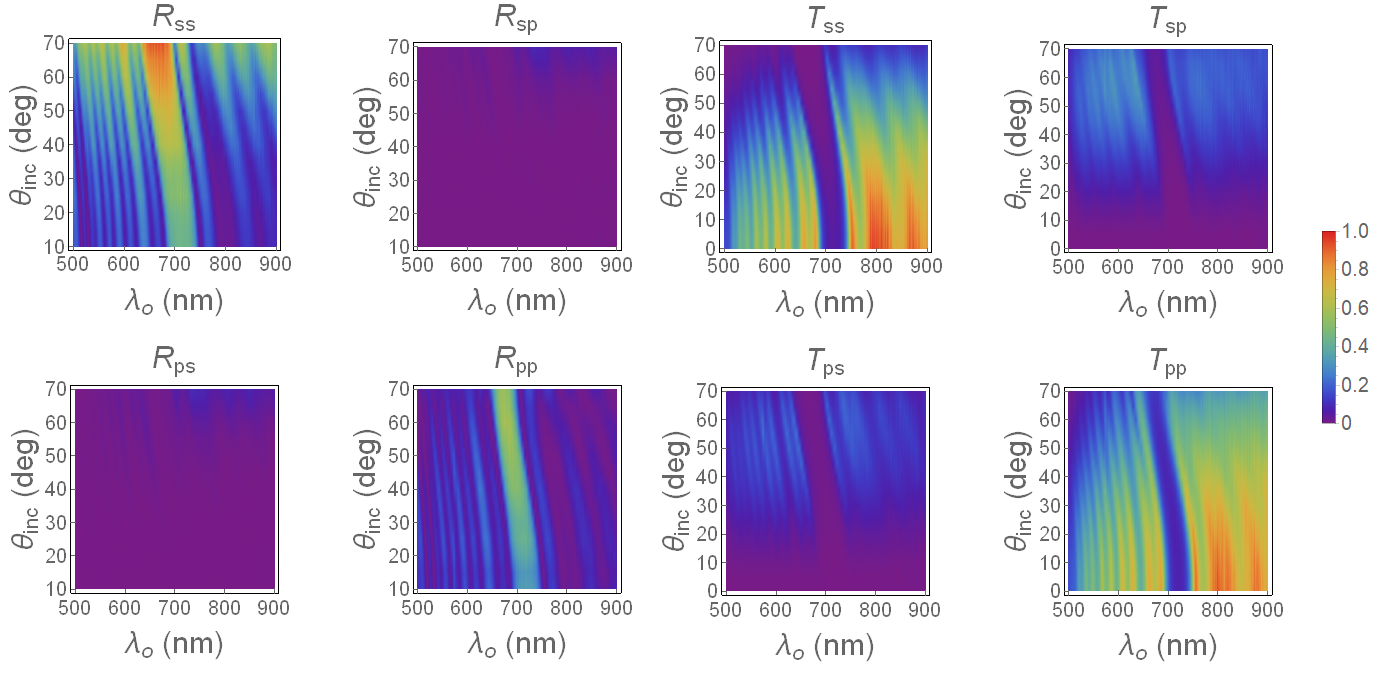}
         \caption{Same as Fig.~\ref{Fig9}, except that
          $\psi=90^\circ$. 
         \label{Fig10}}
\end{figure}

Figure~\ref{Fig10} presents the measured remittances of Sample A  
when $\psi=90^\circ$. Again, the Bragg phenomenon of order $m=2$ is evident in
the plots of $\Rss$, $\Rpp$, $\Tss$, and $\Tpp$. The center wavelengths
are  $\lambdaostwoBr=710$~nm and $\lambdaoptwoBr=720$~nm 
for $\thetainc=10^\circ$ are somewhat redshifted with respect to those for
$\psi=0^\circ$ in Fig.~\ref{Fig9}, in contrast to the slight blueshifts suggested
by the theoretical results presented in Figs.~\ref{Fig3} and \ref{Fig4}.

\subsubsection{Chevronic STF}\label{expt-ChevSTF}

\begin{figure}[ht]
        \centering
        \includegraphics[scale=0.35]{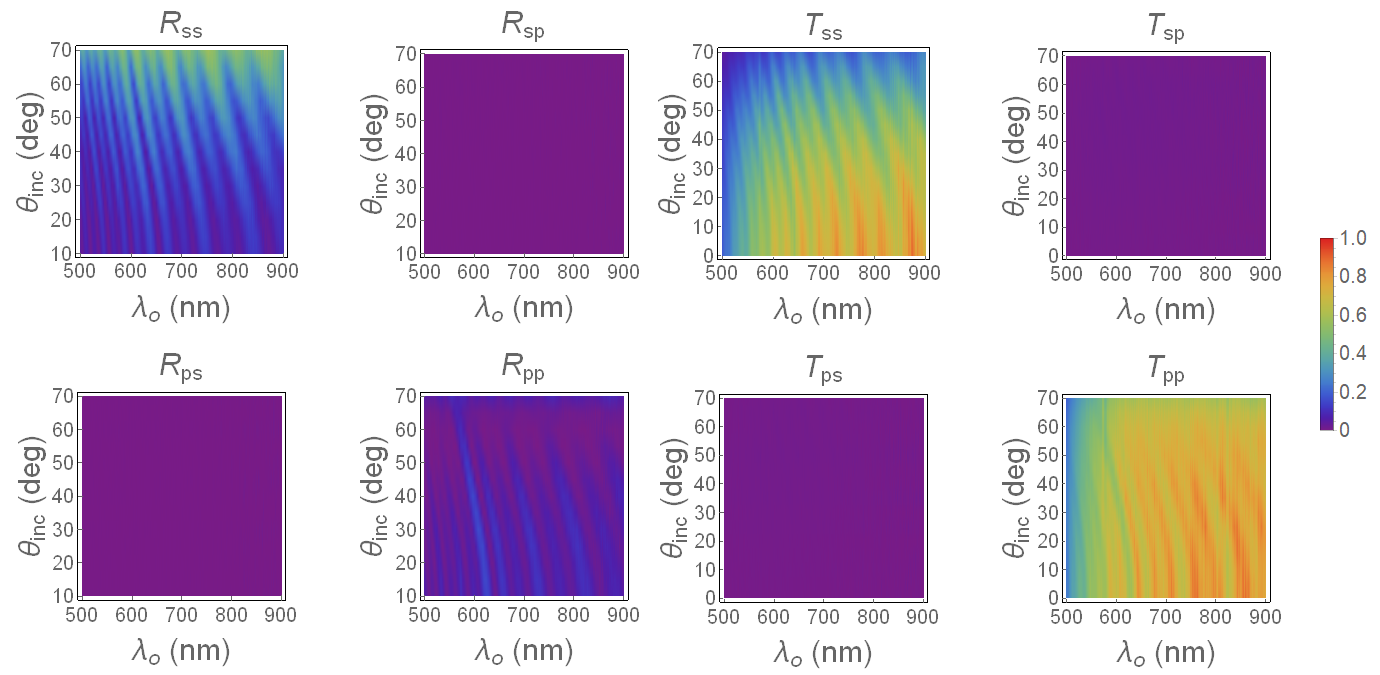}
         \caption{Density plots of all four reflectances and all four transmittances measured as functions
         of $\lambdao$ and $\thetainc$  for Sample B (see Fig.~\ref{Fig2}) when $\psi=0^\circ$.
          \label{Fig11}}
\end{figure}

\begin{figure}[ht]
        \centering
        \includegraphics[scale=0.35]{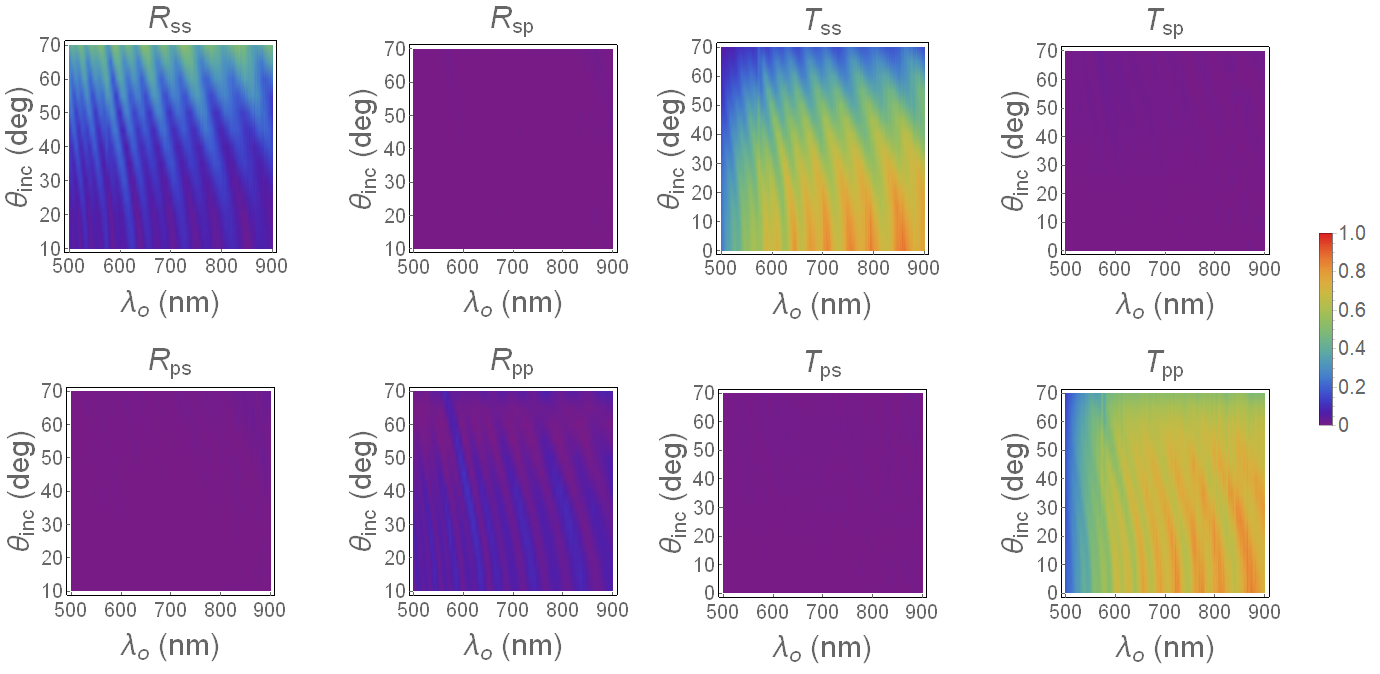}
         \caption{Same as Fig.~\ref{Fig11}, except that
          $\psi=90^\circ$. 
         \label{Fig12}}
\end{figure}

Finally, we present the spectrums of all eight remittances measured for Sample B, which is a ChevSTF. Figure~\ref{Fig11} presents these remittances as functions
of $\lambdao\in[500,900]$~nm and $\thetainc\leq70^\circ$ when $\psi=0^\circ$, and Fig.~\ref{Fig12}  for
$\psi=90^\circ$. Consistently with the theoretical predictions illustrated in Figs.~\ref{Fig5}
and \ref{Fig6}, no evidence of the Bragg phenomenon exists for
$\psi=0^\circ$ in Fig.~\ref{Fig9}. Even the vestigial Bragg phenomenon of order $m=2$ predicted 
in Figs.~\ref{Fig7}
and \ref{Fig8} for highly oblique incidence and $\psi=90^\circ$ is absent in Fig.~\ref{Fig12}.
We could not measure the remittances for $\thetainc>70^\circ$ in order to experimentally investigate the
vestigial Bragg phenomenon of order $m=1$, due to the limitations of our apparatus.

\section{Concluding Remarks}\label{cr}

The unit cell of a chevronic sculptured thin film  comprises two identical columnar thin films (CTFs) except that the nanocolumns of the first are oriented at an angle $\chi$ and nanocolumns of the second are oriented at an angle $\pi-\chi$ with respect to the interface of the two CTFs. A CTF can be modeled as a homogeneous biaxial dielectric material for optical purposes \cite{HWbook}. Thus, a ChevSTF is a structurally periodic piecewise nonhomogeneous
material.

Despite its structural periodicity, we determined that a ChevSTF does not exhibit the Bragg phenomenon when the wave vector of the incident plane wave lies wholly in the  plane
formed by the axes of the nanocolumns of the two CTFs (i.e., when $\psi=0^\circ$).
This conclusion was established both theoretically and experimentally, regardless of
the polarization state of the incident plane wave. 

Theory indicated a vestigial manifestation
of the Bragg phenomenon   for highly oblique incidence when
when the wave vector of the incident plane wave is oriented at an angle $\psi\in(0^\circ,180^\circ)$ with respect to the plane
formed by the axes of the nanocolumns of the two CTFs. Over the limited ranges
of the free-space wavelength and the angle of incidence $\thetainc$ of our
apparatus, we did not observe even that vestigial manifestation.
Thus, we conclude that the Bragg phenomenon would not be observed
 if structural periodicity does not lead to appreciable electromagnetic
periodicity.\\


\noindent\textbf{Acknowledgments}.         
 PDM and AL are grateful to the Charles Godfrey Binder Endowment at the Pennsylvania State University for partial financial support for this research.




\begin{thebibliography}{99} 

\bibitem{Bragg1}
W.~H. Bragg and W. L. Bragg,  ``The reflexion of X-rays by crystals," \emph{Proceedings of the Royal Society of London A} \textbf{88}(605), 428--438 (1913).

\bibitem{Bragg2}
W.~H. Bragg,  ``The intensity of reflexion of X~rays by crystals," \emph{Philosophical Magazine (Series 6)} \textbf{27}(161), 881--899 (1914).

\bibitem{Ewald}
P. P. Ewald, ``Zur Begr\"undung der Kristalloptik. Teil II: Theorie der
Reflexion und Brechung," \emph{Annalen der Physik} \textbf{49}(2), 117--143 (1916).

\bibitem{Ohtaka}
K. Ohtaka, ``Energy band of photons and low-energy photon diffraction,"
\emph{Physical Review B} \textbf{19}(10), 5057--5067 (1979).

\bibitem{Yab}
 E. Yablonovitch, T.~J. Gmitter, and K.~M. Leung, ``Photonic band structure: the face-centered-cubic case employing nonspherical atoms," \emph{Physical Review Letters} \textbf{67}(17), 2295--2298 (1991).

\bibitem{Maldovan}
M. Maldovan and E. L. Thomas,
\emph{Periodic Materials and Interference Lithography},  Wiley--VCH, Weinheim, Germany (2009).

\bibitem {Rayleigh} Lord Rayleigh, ``On the remarkable phenomenon of crystalline reflexion described by Prof. Stokes,"  \emph{Philosophical Magazine (Series 5)} \textbf{26}(160), 256--265 (1888).

\bibitem{HWbook}
I. J. Hodgkinson and Q. h. Wu,
\emph{Birefringent Thin Films and Polarizing Elements},  World Scientific, Singapore (1997).

\bibitem{Macleod}
H. A. Macleod, \emph{Thin-Film Optical Filters, 2nd ed.}, Institute of
Physics, Bristol, United Kingdom (2001).

\bibitem{Mattox}
D.M. Mattox, \emph{The Foundations of Vacuum Coating Technology}, Noyes Publications,  Norwich, NY, USA, 2003.


\bibitem{Baumeister}
P. W. Baumeister,
\emph{Optical Coating Technology},  SPIE Press, Bellingham, WA, USA (2004).


\bibitem{HWapl}
I. Hodgkinson and Q. H. Wu, ``Birefringent thin-film polarizers for use
at normal incidence and with planar technologies," \emph{Applied Physics Letters} {\bf 74}(13),
1794--1796 (1999).

\bibitem{Chen}
H. C. Chen,
\emph{Theory of Electromagnetic Waves},  McGraw--Hill, New York, NY, USA (1983).

\bibitem{Gribble}
C. D. Gribble and A. J. Hall, \emph{Optical Mineralogy, Principles and Practice}, University
College London Press, London, United Kingdom (1992).




\bibitem{STFs}
A. Lakhtakia and R. Messier,
\emph{Sculptured Thin Films: Nanoengineered Morphology and Optics},  SPIE Press, Bellingham, WA, USA (2005).


\bibitem{NoBP2}G. Y. Slepyan and A. S. Maksimenko, ``Motohiro-Taga interface in sculptured thin films---absence of Bragg phenomena,'' \emph{Optical Engineering} \textbf{37}(10), 2843--2847 (1998).

\bibitem{NoBP1} A. Lakhtakia and R. Messier, ``Reflection at the Motohiro-Taga interface of two anisotropic materials with columnar microstructures,'' \emph{Optical Engineering} \textbf{33}(8), 2529--2534 (1994).

\bibitem{CBP} 
M. Faryad and A. Lakhtakia, ``The circular Bragg phenomenon,'' \emph{Advances in Optics and Photonics} \textbf{6}(2), 225--292 (2014).


\bibitem{SLml}
S. E. Swiontek and A. Lakhtakia, ``Vacuum-metal-deposition and columnar-thin-film techniques
implemented in the same apparatus," \emph{Materials Letters} \textbf{142}(1), 291--293 (2015).

\bibitem{transmission_1} 
{[https://www.thorlabs.com/NewGroupPage9.cfm?ObjectGroup$_-$ID=3981]} (Februrary 22, 2017).

\bibitem{transmission_2} 
{[http://www.tydexoptics.com/materials1/for$_-$transmission$_-$optics/cvd$_-$znse/]}
(accessed Februrary 22, 2017).


\bibitem{Alfa} {[https://www.alfa.com/en/catalog/013241/]}
(accessed Februrary 22, 2017).

\bibitem{Biaxial3} I. J. Hodgkinson, Q. h. Wu, and J. Hazel, ``Empirical equations for the principal refractive indices and column angle of obliquely deposited films of tantalum oxide, titanium oxide, and zirconium oxide,'' \emph{Applied Optics} \textbf{37}(13), 2653--2659 (1998).

\bibitem{Erten}
S. Erten, A. Lakhtakia, and G. D. Barber, ``Experimental investigation of circular Bragg phenomenon for oblique incidence," \emph{Journal of the Optical Society of America A} {\bf 32}(5), 764--770 (2015).





\bibitem{Freeze_Fracture} 
N. J. Severs, ``Freeze-fracture electron microscopy,'' \emph{Nature Protocols} \textbf{2}(3), 547--576 (2007). 

\bibitem{Messier97}
R. Messier, T. Gehrke, C. Frankel, V. C. Venogopal, W. Ota{\~n}o, and A. Lakhtakia,
``Engineered sculptured nematic thin films," \emph{Journal of Vacuum Science and Technology A} {\bf 15}(4), 2148--2152 (1997).

\bibitem{Marple}
D. T. F. Marple,
``Refractive index of ZnSe, ZnTe, and CdTe," \emph{Journal of Applied Physics} {\bf 35}(3), 539--542 (1964).


\bibitem{Devore}
J. R. Devore, ``Refractive indices of rutile and sphalerite," \emph{Journal of the Optical Society of America} {\bf 41}(6), 416--419 (1951).

\bibitem{StJohn}
W. D. St. John, W. J. Fritz, Z. J. Lu, and D.-K. Yang, ``Bragg reflection
from cholesteric liquid crystals," \emph{Physical Review E} {\bf 51}, 1191--1198 (1995).






\end{thebibliography}
\end{document}